\documentclass[showpacs,preprintnumbers,amsmath,amssymb,superscriptaddress,nofootinbib,english]{revtex4}

\usepackage{graphicx}
\usepackage{dcolumn}
\usepackage{bm}
\usepackage{epsfig}
\usepackage{graphicx}
\usepackage{hyperref}
\usepackage[usenames]{color}
\usepackage{url}

\hypersetup{
    colorlinks=true,
    linkcolor=green,
    citecolor=blue,
}

\newcommand{\remove}[1]{}

\def\be{\begin{equation}}
\def\ee{\end{equation}}
\def\ba{\begin{eqnarray}}
\def\ea{\end{eqnarray}}

\begin{document}

\title{ Modified Gravity and the CMB}

\author{Philippe~Brax}
\email[Email address: ]{philippe.brax@cea.fr}
\affiliation{Institut de Physique Theorique, CEA, IPhT, CNRS, URA 2306, F-91191Gif/Yvette Cedex, France}

\author{Anne-Christine~Davis}
\email[Email address: ]{a.c.davis@damtp.cam.ac.uk}
\affiliation{DAMTP, Centre for Mathematical Sciences, University of Cambridge, Wilberforce Road, Cambridge CB3 0WA, UK}

\date{\today}

\begin{abstract}
We consider the effect of modified gravity on the peak structure of the Cosmic Microwave Background (CMB) spectrum. We focus on simple models of modified gravity mediated by a massive scalar field coupled to both baryons and cold dark matter. This captures the features of chameleon, symmetron, dilaton and $f(R)$ models. We find that the CMB peaks can be affected in three independent ways provided the Compton radius of the massive scalar is not far-off the sound horizon at last scattering. When the coupling of the massive scalar to Cold Dark Matter (CDM) is large, the anomalous growth of the CDM perturbation inside the Compton radius induces a change in the peak amplitudes. When the coupling to baryons is moderately large, the speed of sound is modified and the peaks shifted to higher momenta. Finally when both couplings are non-vanishing, a new contribution proportional to  the Newton potential appears in the Sachs-Wolfe temperature and increases the peak amplitudes. We also show how, given any temporal evolution of  the scalar field mass, one can engineer a
corresponding modified gravity model of the chameleon type. This opens up the possibility of having independent constraints on modified gravity from the CMB peaks and large scale structures at low redshifts.
\end{abstract}

\maketitle

\section{Introduction}

Modified gravity is a tantalising idea which has received a lot of attention recently, see \cite{Jain:2010ka,Clifton:2011jh} and references therein. This is partially due to the quest
for an explanation to the acceleration of the Universe. Another reason is the possibility of constructing meaningful models of massive gravity.
In most of these attempts to modify gravity, scalar fields play a prominent role. From the dark energy point of view, scalar fields
are plagued with the fifth force problem whereby the scalars generating the acceleration of the Universe would lead to an
extraneous contribution to Newton's law. This can be cured by screening mechanisms locally. Indeed, dense regions such as the sun or the earth
can trap the scalar field and a minute deviation to Newton's law ensues. This the case of the chameleon models \cite{Khoury:2003aq,Khoury:2003rn,Brax:2004qh,Mota:2006ed,Mota:2006fz}. In other cases such as the galileon \cite{Nicolis:2008}, symmetron \cite{Hinterbichler:2010es,Olive:2007aj} and dilaton \cite{Brax:2010gi}, the coupling to matter almost vanishes in dense environments.

Unfortunately, these models behave like $\Lambda$-CDM at the background level at least since Big Bang Nucleosynthesis jeopardising any attempt to distinguish them
from the cosmological standard model. Fortunately, they modify gravity at the perturbation level and lead to observable effects on the growth of structures in the recent past of the Universe. Typically, gravity is modified inside the Compton radius of the scalar field which must be of order of a few Mpc now. As the growth of structures is observed using galaxies as tracers of Newton's potential, the analysis  is hampered by the galaxy bias and therefore subject to some uncertainty.
On the other hand, the physics of the CMB is linear and evades all non-linear problems. The CMB could serve as a clean template for the seach
of modified gravity effects. This is what we will pursue in this paper.

Modified gravity will be most effective at last scattering when the Compton radius is close to the sound horizon. In this case, scales inside the sound horizon
would be affected by a modification of gravity. An extreme case, where the mass vanishes and all scales are affected by modified gravity, has been considered in the
Brans-Dicke framework in \cite{Chen:1999qh,Acquaviva:2004ti} and recently revisited by  \cite{Ali:2010gr} where it has been  found that the constraints are less severe than the Cassini bound on the Eddington parameter \cite{Bertotti:2003rm}.
Here we will always consider the massive case with a Compton radius in the range of the CMB peak scales. In this case, three effects can occur. The first one is driven by the coupling to CDM and consists in an anomalous growth of Newton's potential. This increases the peak amplitude. Another effect results from a decrease of the speed of sound coming from the coupling to baryons. If this coupling is too large, an instability with an imaginary speed of sound sets in as noticed in \cite{Bean:2007ny}. Finally a more subtle effect comes from both couplings and induces a modification of the contribution of Newton's potential to the Sachs-Wolfe effect \cite{Rhodes:2003ev,Mainini:2010ng}.

We also show that the  knowledge of the scalar field mass around the last scattering surface is enough to reconstruct some part of the interaction potential of a chameleon model with such a mass and  couplings. If some knowledge of the mass function at later times could also be inferred from large scale structures, a larger
range of the interaction potential could be rebuilt. Eventually, if the mass function could also be extracted from data in the intermediate regime corresponding to the cosmic dark ages, an almost complete picture of the interaction potential of the scalar field could be drawn. Together with the values of the couplings to baryons and CDM which could be obtained from the deviations of the CMB peaks from their $\Lambda$-CDM values, an almost complete description of the source of modified gravity since
the last scattering epoch could be obtained.

In the first part, we analyse linear perturbations and the growth of structures in a modified gravity context. Then we use the tight binding approximation \cite{Hu:1994uz}
to study the effects of modified gravity on the CMB peaks. Then, we show that one can always engineer any modification of gravity on large scales by
choosing the time evolution of the scalar field mass. Finally we discuss our
conclusions.

\section{Modified Gravity and Linear perturbations}

\subsection{Perturbations}

Modified gravity will be modelled using a massive scalar field $\phi$ with a time varying mass $m(a)$ and coupling constants $\beta_b$ and $\beta_c$ to both baryons and Cold Dark Matter (CDM). We will not assume that the couplings are constant, as some interesting models such as the environmentally dependent dilaton have field dependent couplings \cite{Brax:2010gi}. On large cosmological scales and at the perturbative level, models of modified gravity such as $f(R)$ theories, chameleons, dilatons, symmetrons and galileons all involve a scalar field with  couplings to matter \cite{
Khoury:2003rn,Mota:2006fz,Hinterbichler:2010es,Brax:2010gi,Nicolis:2008,Olive:2007aj}.  On shorter distances, one must invoke the existence of a screening mechanism in order to evade tight gravitational tests in the solar system, such as either the chameleon or the Vainshtein screening properties. At the level of the CMB, the screening of gravity on small scales is not present and large effects may ensue, depending on the range  of the fifth force mediated by the scalar field and the intensity of the coupling to matter.

We focus on scalar fields in the Einstein frame where the Einstein equations are preserved and depend on the energy momentum tensor of the scalar field:
\begin{equation}
T^\phi_{\mu\nu}= \partial_\mu\phi \partial_\nu \phi -g_{\mu\nu} ( \frac{1}{2} (\partial \phi)^2 +V).
\end{equation}
Matter couples to both gravity and the scalar field via the metrics
\begin{equation}
g^{i}_{\mu\nu} =A^2_i(\phi) g_{\mu\nu}, \ \ i=b,c
\end{equation}
for each matter species. Their couplings to matter are defined as
\begin{equation}
\beta_i(\phi)= m_{\rm Pl}\frac{\partial \ln A_i(\phi)}{\partial \phi}
\end{equation}
which may be field dependent in dilatonic models for instance. It is a universal constant $\beta=\frac{1}{\sqrt 6}$ in $f(R)$ models.

In all the models that we will consider, the background cosmology is tantamount to a $\Lambda$-CDM model since Big Bang Nucleosynthesis (BBN). Deviations from General Relativity only  appear at the perturbative level.
Moreover, the effective potential for the scalar field in the matter era is modified by the presence of matter as a consequence of the non-trivial matter couplings
\begin{equation}
V_{\rm eff}(\phi)= V(\phi) + \sum_i A_i(\phi) \rho_i
\end{equation}
where the sum is taken over the non-relativistic species and $\rho_i$ is the conserved energy density of the ith fluid. The potential acquires
a slowly varying minimum $\phi(\rho_i)$ which is an attractor as long as the mass $m^2_{\rho}= \frac{d^2 V_{eff}}{d^2\phi}\vert_{\phi(\rho_i)}$ is larger than the Hubble rate
\cite{Brax:2004qh}. We will always assume that this is the case in the following.
Due to the interaction with the scalar field, matter is not conserved and satisfies
\begin{equation}
D_\mu T^{\mu\nu}_i= \kappa_4\beta_i \partial^\mu \phi T^i.
\end{equation}
where $T^{\mu\nu}_i$ is the energy momentum tensor of the ith species, $T$ its trace and $\kappa_4^2 =8\pi G_N$. Notice that radiation is not affected by the presence of the coupled scalar field.
The Einstein equation is not modified and reads
\begin{equation}
R_{\mu\nu}-\frac{1}{2} g_{\mu\nu} R= 8\pi G_N (\sum_i T^i_{\mu\nu} +T^\phi_{\mu\nu})
\end{equation}
where $m_{\rm Pl}^{-2}\equiv 8\pi G_N$.
The conservation equations and the Einstein equations are enough to characterise the time evolution of the fields.

We are interested in the first order perturbations of the Einstein equation and the conservation equations. In the absence of anisotropic stress, the metric can be described in the conformal Newton gauge
\begin{equation}
ds^2= a(\eta)(-(1+2\Phi) d\eta^2 + (1-2\Phi) dx^2)
\end{equation}
where $\Phi$ is Newton's potential. In this gauge the relevant Einstein equations involve $\delta T_i^0$ and $\delta T^0_0$. For the scalar field we find that
\begin{equation}
\delta T_i^0= -\frac{1}{a^2} \phi' \partial \delta \phi
\end{equation}
where $'=d/d\eta$ and
\begin{equation}
\delta T^0_0= \frac{\phi'^2}{a^2} \Phi -\frac{\phi'}{a^2} \delta \phi' -V_\phi \delta \phi
\end{equation}
which implies that the Poisson equation becomes (in Fourier modes)
\begin{equation}
k^2\Phi + 3 {\cal H}( \Phi' + {\cal H} \Phi)= -4\pi G_N a^2 \sum_i \delta \rho_i -4\pi G_N a^2 (\frac{\phi'^2}{a^2} \Phi -\frac{\phi'}{a^2} \delta \phi' -V_\phi \delta \phi)
\end{equation}
and
\begin{equation}
k^2 (\Phi' + {\cal H} \Phi)= 4\pi G_Na^2 \sum_i (1+ w_i) \theta_i + 4\pi G_N k^2 \phi' \delta \phi
\end{equation}
where the Hubble rate is ${\cal H}= a'/a$ and the equation of state of each species is $w_i$. We have defined the divergence of the velocity field of each species $\theta= \partial_i v^i$.
The perturbed Klein-Gordon equation can be expressed as
\begin{equation}
 \delta\phi'' +2{\cal H} \delta \phi' +(k^2 + a^2 m^2)\delta \phi + 2\Phi a^2  V_{\rm eff\phi}-4\Phi' \phi'+ \kappa_4 a^2 (\beta_c A_c\delta \rho_c +\beta_b A_b\delta \rho_b)=0
 \end{equation}
where  $\delta \rho_i =\rho_i \delta_i$.
We have denoted by $\rho_b$ the baryon energy density and $\rho_c$ the CDM energy density.

The conservation equation for the CDM particles lead to the coupled equations
\begin{equation}
\delta_c' = -(\theta_c -3 \Phi') + \kappa_4 \beta_\phi^c \phi' \delta\phi + \kappa_4 \beta_c A_c\delta\phi'
\end{equation}
and
\begin{equation}
 \theta_c'= -{\cal H} \theta_c + k^2 \Phi +\kappa_4 k^2 \beta_c A_c \delta \phi -\beta_c A_c \phi' \theta_c.
\end{equation}
The baryons are coupled to the photons by Thompson scattering implying that the conservation equations
are modified
\begin{equation}
 \delta_b'=-\theta_b +3 \Phi' +\kappa_4 \beta^b_\phi \phi' \delta \phi + \beta_b \kappa_4 A_b \delta\phi'
\end{equation}
and
\begin{equation}
 \theta_b'= -{\cal H}\theta_b +\frac{an_e \sigma_T}{R} (\theta_\gamma-\theta_b)+ k^2\Phi + \beta_b A_b\kappa_4 k^2 \delta \phi -\beta_b A_b \phi' \theta_b.
\end{equation}
The Thompson scattering cross section depends on the fine structure constant $\alpha$ and the electron mass $m_e$. The electron mass is scalar field dependent due to the conformal rescaling of the metric, i.e. $m_e$ is proportional to $A_b(\phi)$. Perturbations of the scalar field would have an effect at the level of second order perturbation theory. At the linear level, the dependence of the electron mass on $\phi$ must only be taken into account at the background level. As the scalar field tracks the minimum of the effective potential, the time variation of the mass is electron mass in one Hubble time is suppressed by ${\cal O} (\frac{H^2}{m^2}) \ll 1$, see (\ref{evo}), and can therefore be neglected. If the fine structure constant depends on the scalar field, its time dependence can also be neglected for the same reason.

We will identify the speed of sound in the absence of modified gravity as
\begin{equation}
 c_s^2= \frac{1}{3} \frac{1}{1+R}
\end{equation}
where the baryon to photon ratio is
\begin{equation}
R= \frac{3}{4} \frac{\rho_b}{\rho_\gamma}
\end{equation}
which is about $0.6$ at last scattering.
Finally we shall work in the fluid approximation for the photons. The Boltzmann hierarchy is not altered by the presence of a scalar field.
We find
\begin{equation}
\delta_\gamma'= -\frac{4}{3} \theta_\gamma
\end{equation}
and
\begin{equation}
 \theta_\gamma'= \frac{k^2}{4} \delta_\gamma + k^2 \Phi + an_e \sigma_T (\theta_b-\theta_\gamma).
\end{equation}
This completes our description
of the perturbation equations.

We also need to specify the initial conditions for all the perturbations. We will be interested in modes which will enter the horizon before radiation-matter equality. Adiabatic initial conditions are determined by
\begin{equation}
\delta_c^0= \delta_b^0= \frac{3}{4} \delta_\gamma^0.
\end{equation}
This is related to the initial Newton potential as
\begin{equation}
\delta_b^0= -\frac{3}{2} \Phi_0.
\end{equation}
We will always express the perturbed quantities in terms of the initial Newton potential.

\subsection{Growth of structures}

The growth of CDM perturbations is affected by the modification of gravity due to the scalar field. Let us consider sub-horizon perturbations where $k\gg {\cal H}$. We assume that the background configuration of the scalar field tracks the minimum of the effective potential:
\begin{equation}
V_\phi=-\kappa_4 \beta_c A_c\rho_c -\kappa_4 \beta_b A_b\rho_b.
\end{equation}
This implies that we can also neglect terms in $\phi'$ as
\begin{equation}
\frac{\kappa_4 \phi'}{H}= 9\sum_i A_i \beta_i \Omega_i \frac{H^2}{ m^2}
\label{evo}
\end{equation}
where $m\gg H$.
In the sub-horizon limit, we also neglect the time variation of Newton's potential as terms in $k^2\Phi$ dominate over $\dot \Phi$.
In this case the Klein-Gordon equation becomes an algebraic equation
\begin{equation}
\delta \phi= -\kappa_4 \frac{\beta_c A_c \delta\rho_c + \beta_b A_b \delta \rho_b}{\frac{k^2}{a^2} +m^2}.
\label{dp}
\end{equation}
As $\beta_b$ can differ from $\beta_c$, the baryon contribution to the scalar field perturbation cannot be neglected.
Similarly the Poisson equation becomes
\begin{equation}
k^2 \Phi=- 4\pi G_N a^2 (\delta\rho_c +\delta\rho_b +\delta\rho_\gamma).
\label{poi}
\end{equation}
We can neglect the contribution of the baryon and photon  perturbations $\delta_{b,\gamma}= \frac{\delta \rho_{b,\gamma}}{\rho_{b,\gamma}}$ to Newton's potential in the matter era as the CDM density contrast $\delta_c=\frac{\delta \rho_c}{\rho_c}$ grows rapidly. Similarly in the radiation era, the baryons and photons are tightly coupled and their oscillatory perturbations average to zero. As a consequence we can neglect the baryons and photons in the Poisson equation up to the last scattering time.
The CDM equations become
\begin{equation}
\delta_c' = -\theta_c
\end{equation}
and
\begin{equation}
\theta_c'= -{\cal H} \theta_c + k^2 \Phi + \beta_c A_c  k^2 \kappa_4 \delta\phi
\end{equation}
leading to the growth equation
\begin{equation}
\delta_c'' +{\cal H} \delta_c' -\frac{3}{2} {\cal H}^2 \frac{\rho_c}{\rho_c +\rho_\gamma +\rho_b} (1+ \frac{2\beta_c^2 A_c^2}{1+\frac{m^2a^2}{k^2}})=0
\end{equation}
When $A_c \sim 1$ and $\rho_c\gg \rho_b,\rho_\gamma$, this is an equation which can be easily integrated deep inside the Compton radius $k/a \gg m$ and outside the Compton radius $k/a \ll m$ in the matter dominated era. Outside the Compton radius, the CDM density contrast grows like:
\begin{equation}
\delta_c \sim a
\end{equation}
as in General Relativity while inside the Compton radius the
growth of structures is enhanced due to the factor $(1+ 2\beta_c^2)$ corresponding to an increase of Newton's constant
\begin{equation}
\delta_c \sim a^{\frac{-1+\sqrt{1+24(1+2\beta_c^2)}}{4}}
\end{equation}
In particular this implies that Newton's potential is not constant in time on small enough scales. A well known consequence of this fact is a new contribution to the Integrated Sachs Wolf (ISW) effect for the CMB spectrum, see \cite{Rhodes:2003ev} and references therein.
Finally, during the radiation era the baryonic and photonic perturbations are tightly coupled and oscillates while the CDM perturbations grow logarithmically as
\begin{equation}
\delta_c= \delta_0 (1+ 2 \frac{{\cal H}_k}{k_{eq}} \frac{a_k}{a_{eq}} \ln \frac{a}{a_k})
\end{equation}
where $a_k$ and ${\cal H}_k$ are the time when the momentum $k$ enters the horizon, when ${\cal H}_k=k$,  and $k_{\rm eq}$ is such that scales with $k>k_{eq}$ enter
the horizon before matter-radiation equality.
It turns out that
$\frac{{\cal H}_k}{k_{eq}} a_k=a_{eq}$ implying that
\begin{equation}
\delta_c= \delta_0 (1+ 2 \ln \frac{a}{a_k})
\end{equation}
where $\delta_0$ is the density contrast at horizon entry. Initially we have that $\delta_0=-\frac{3}{2}\Phi(0)$ where $\Phi (0)$ is the constant Newton potential
outside the horizon and we have taken $\dot \Phi\vert_{a_k}=0$ which implies that $\frac{d \ln \delta _c}{d\ln a}\vert_{a_k}=2$.

The linear approximation can be used provided the screening of the scalar field effects is not present. The thin-shell mechanism occurs when the variation of the scalar field profile due to an overdensity satisfies
\begin{equation}
\kappa_4 \delta \phi \ll 6\beta_c \Phi
\end{equation}
where we simplify the problem and assume  that only CDM is responsible for the overdensity. Using (\ref{dp}) and (\ref{poi}), this is equivalent to
\begin{equation}
k\ll am(a).
\end{equation}
In the following we will be interested in scales which are close enough to the Compton wavelength, implying that the linear analysis is valid.

At the beginning of the matter era, the growth of perturbations depends on the behaviour of $a m(a)$. When $a m(a)$ increases, the scale $k$  can already be inside the Compton radius at matter equality and then grow anomalously before leaving the Compton radius. Schematically, this leads to a growth in $\ln a$, followed by $a^{\nu/2}$, followed by $a$. If the scale $k$ is not in the Compton radius at equality and $am(a)$ increases, it will never be. On the other hand when $am(a)$ decreases, if the scale $k$ is not in the Compton radius at equality, it will enter the Compton radius at a later time implying a growth in $\ln a$, followed by $a$, and then $a^{\nu/2}$.
These patterns will be explicit in the examples we will present in the next section.

\subsection{From horizon entry to last scattering}

We will give an explicit example of anomalous growth of structures from horizon entry to the last scattering surface. This will make explicit the general remarks of the
previous section and will be useful when we discuss the CMB in the tight coupling approximation \cite{Hu:1994uz}.
We focus on models where the time dependence of the mass is given by
\begin{equation}
m(a)= m_1(\frac{a}{a_{LS}})^r
\end{equation}
where $a_{LS}$ is the scale factor at last scattering and $r$ is an index. The scale $m_1$ will be chosen to be of the order of the (inverse) CMB length scales, i.e. close to the sound horizon (see sections III-C and III-D for a description of models leading to this power law behaviour ).
The comoving Compton length increases when $r>-1$ and decreases when $r<-1$. This implies that scales enter the Compton radius when $r<-1$, and they eventually leave the Compton radius when $r>-1$. More precisely,
scales coincide with   the Compton radius when
\begin{equation}
a_{\rm Compton}= a_{LS} (\frac{k}{k_1})^{1/(r+1)}
\end{equation}
where $k_1= a_{LS} m_1$. When $r<-1$, scales $k<k_1$ enter the Compton radius after last scattering while scales $k>k_1$ enter the Compton length before last scattering. When $r>-1$, scales with $k>k_1$ leave the Compton radius after last scattering while scales $k<k_1$ leave the Compton radius before last scattering.

Let us consider a given scale $k$ which enters the horizon before matter-radiation equality. Up to matter-radiation equality, the density contrast $\delta_c$ grows logarithmically. After the equivalence, one must distinguish whether $am(a)$ increases or decreases.

\subsubsection{$am(a)$ decreases}

This corresponds to $r<-1$. It is very useful to determine when scales enter the Compton horizon after matter-radiation equality. This happens when
\begin{equation}
k<k_c=k_1(\frac{1+z_{eq}}{1+z_{LS}})^{-(r+1)}
\end{equation}
Hence when $k_{eq}<k<k_1$, scales enter the Compton length after last scattering. Here  $k_{eq}$ is such that modes $k>k_{eq}$ enter the horizon before matter-radiation equality. Such a mode grows logarithmically till matter-radiation equality and then undergoes a normal growth implying that
\begin{equation}
\frac{\delta_{LS}}{\delta_0}=\frac{a_{LS}}{a_{eq}}(1+ 2 \ln \frac{a_{eq}}{a_k})
\end{equation}
When $k_1<k<k_c$, a mode first grows logarithmically till
matter-radiation equality, then grows at a normal rate when outside the Compton radius and then grows anomalously.
The density contrast at Compton radius entry is
\begin{equation}
\frac{\delta_c}{\delta_0}= \frac{a_{\rm Compton}}{a_{eq}}(1+ 2 \ln \frac{a_{eq}}{a_k}).
\end{equation}
The growth is anomalous between $a_{\rm Compton}$ and $a_{LS}$ implying that
\begin{equation}
\frac{\delta_{LS}}{\delta_0}=\frac{a_{LS}}{a_{eq}} (\frac{k}{k_1})^{(1-\nu/2)/(r+1)}(1+ 2 \ln \frac{a_{eq}}{a_k}).
\end{equation}
Notice the anomalous power law dependence.
When $k>k_c$, the density contrast is simply
\begin{equation}
\frac{\delta_{LS}}{\delta_0}=(\frac{a_{LS}}{a_{eq}})^{\nu/2}(1+ 2 \ln \frac{a_{eq}}{a_k}).
\end{equation}
which corresponds to an anomalously enhanced density contrast compared to General Relativity.

\subsubsection{$am(a)$ increases}

This corresponds to $r>-1$. Scales leave the Compton horizon after matter-radiation equality when
\begin{equation}
k>k_c=k_1(\frac{1+z_{LS}}{1+z_{eq}})^{r+1}
\end{equation}
Hence when $k_{eq}<k<k_c$,  a mode first grows logarithmically till
matter-radiation equality and then grows at a normal rate  until last scattering as it is outside the Compton radius.
On the other hand, when $k_1>k>k_c$ modes grow logarithmically,  then anomalously and eventually at a normal rate when they have left the Compton radius before last scattering. The density contrast when leaving the  Compton length is
\begin{equation}
\frac{\delta_c}{\delta_0}= (\frac{a_{\rm Compton}}{a_{eq}})^{\nu/2}(1+ 2 \ln \frac{a_{\rm eq}}{a_k}).
\end{equation}
The growth is normal  between $a_{\rm Compton}$ and $a_{LS}$ implying that
\begin{equation}
\frac{\delta_{LS}}{\delta_0}=\frac{a_{LS}}{a_{eq}} (\frac{k}{k_1})^{(\nu/2-1 )/(r+1)}(1+ 2 \ln \frac{a_{eq}}{a_k}).
\end{equation}
Notice the anomalous power law dependence. Finally when $k>k_1$, scales grow logarithmically and then anomalously
\begin{equation}
\frac{\delta_{LS}}{\delta_0}=(\frac{a_{LS}}{a_{eq}})^{\nu/2}(1+ 2 \ln \frac{a_{eq}}{a_k})
\end{equation}
which corresponds to an anomalously enhanced density contrast compared to General Relativity.

\subsubsection{Discussion}
We have deduced that the growth of the CDM density contrast is affected in an interval due to the fact that
scales enter/leave the Compton radius. When such an anomalous power law is present, we find therefore that Newton's potential is not
growing logarithmically only but
\begin{equation}
k^2 \Phi \propto (\frac{k}{k_1})^{(\nu/2-1 )/(r+1)}(1+ 2 \ln \frac{a_{eq}}{a_k})
\end{equation}
which is due to the modification of gravity in the $[k_1,k_c]$ (respectively $[k_c,k_1]$) interval.
In the numerical examples, we will take $r=1$ and $r=-2$.

\section{ The Cosmic Microwave Background}

\subsection{The tight binding approximation}

We are interested in the temperature fluctuations of the CMB as given by the Sachs-Wolfe formula
\begin{equation}
\frac{\delta T}{T}= \frac{1}{4} \delta _\gamma + \Phi + e_i v^i_b + 2 \int_E^0 \Phi' d\eta
\end{equation}
where $e_i$ is a vector along the line of sight.
Effects of modified gravity are multiple. As we have already mentioned, the fact that Newton's potential varies in time leads to a contribution to the ISW effect, see for instance  \cite{Rhodes:2003ev} and references therein.
Here we are interested in intermediate scales where the peak structure of the CMB is relevant. We will study the effect of modified gravity on the temperature anisotropy
\begin{equation}
\Theta= \frac{1}{4} \delta _\gamma + \Phi
\end{equation}
using the tight binding approximation \cite{Hu:1994uz}, allowing us to capture the essence of the consequences of modified gravity on the CMB peaks. This approach is not precise enough to impose tight bounds on the couplings. We will only use it to analyse the possible consequences of modified gravity on the CMB qualitatively.

On sub-horizon scales and neglecting the time variation of $\phi$ we have
\begin{equation}
\delta_b'= - \theta_b
\end{equation}
and
\begin{equation}
\delta_\gamma'= -\frac{4}{3} \theta_\gamma.
\end{equation}
In the tight binding approximation, the photon and baryon density contrasts are linked by
\begin{equation}
an_e \sigma_T (\theta_b-\theta_\gamma)= \theta_\gamma' -\frac{k^2}{4} \delta_\gamma -k^2 \Phi
\end{equation}
which does not involve the scalar field directly. When $an_e \sigma_T$ is larger than the Hubble rate this implies that
\begin{equation}
\theta_b\approx \theta_\gamma
\end{equation}
and therefore
\begin{equation}
\delta_b\approx \frac{3}{4}\delta_\gamma
\end{equation}
leading to
\begin{equation}
\delta_b'' + \frac{R'}{1+R} \delta_b' + c_s^2 k^2 \delta_b=-k^2 \Phi - \kappa_4 \beta_b A_b k^2 \frac{R}{R+1} \delta\phi.
\end{equation}
It is particularly useful to define
\begin{equation}
\delta_b =(1+R)^{-1/2} \delta
\end{equation}
from which we deduce that
\begin{equation}
\delta'' + c_s^2 k^2 \delta=-k^2(1+R)^{1/2}\Phi - \kappa_4 \beta_b A_b k^2 \frac{R}{(R+1)^{1/2}} \delta\phi
\end{equation}
in the sub-horizon limit.
Neglecting the baryons and photons in the Poisson equation we find that
\begin{equation}
\delta'' + c_s^2 k^2 (1- \frac{9\Omega_b \beta_b^2 A_b^2 R{\cal H}^2 }{{k^2} +m^2a^2})\delta=-k^2(1+R)^{1/2}(1+ \frac{2\beta_bA_b \beta_c A_b}{1+\frac{m^2a^2}{k^2}} \frac{R}{R+1})\Phi.
\end{equation}
Finally we can introduce
\begin{equation}
\tilde \delta= \delta + (1+R)^{1/2}(1+ \frac{2\beta_bA_b \beta_c A_b}{1+\frac{m^2a^2}{k^2}} \frac{R}{R+1})\frac{\Phi}{\tilde c_s^2}
\end{equation}
where the effective speed of sound is
\begin{equation}
\tilde c_s^2= c_s^2  (1- \frac{9\Omega_b \beta_b^2 A_b^2 R{\cal H}^2 }{{k^2} +m^2a^2})
\end{equation}
which leads to
\begin{equation}
\tilde \delta'' +\tilde c_s^2 \tilde\delta=0.
\end{equation}
The effective speed of sound can become negative when $\beta_b$ is too large leading to an instability already noted in \cite{Bean:2007ny}.
The WKB solutions of this equations are
\begin{equation}
\tilde \delta= \tilde c_s^{-1/2}(A \sin k \tilde  r_s + B \cos k \tilde r_s)
\end{equation}
where $\tilde r_s(\eta)= \int_{\eta_k}^\eta \tilde c_s d\eta$ is the modified sound horizon. We have taken the initial condition to be at $\eta_k$ when
scales becomes sub-horizon, ${\cal H}=k$.
The WKB approximation is valid as long as $w'\ll w^2$ where $w= k\tilde c_s$. This is valid inside the horizon.
Initially both $\delta_\gamma$ and $\Phi$ are constant  implying that
\begin{equation}
\tilde \delta = 3^{1/4} (1+R)^{1/4} B \cos \tilde r_s k
\end{equation}
and we find that
\begin{equation}
B= \frac{(1+R_k)^{1/4}}{3^{1/4}} (1+2R_k) \frac{3}{2}\Phi(0)
\end{equation}
where we have set the initial conditions as adiabatic at horizon entry in the radiation era.
This
leads to
\begin{equation}
\tilde \delta= \frac{3}{2}(1+R_k)^{1/4}(1+2R_k) \Phi(0) (1+R)^{1/4} \cos r_s k
\end{equation}
where we have used $\delta_b(0) =-\frac{3}{2}\Phi (0)$ outside the horizon.
We have also used the fact that initially all scales are outside the horizon and thus outside the Compton radius.
It is important to relate $\tilde\delta$ to $\Theta$
\begin{equation}
\Theta= \frac{\tilde \delta}{3 (1+R)^{1/2}} +(1-3 \tilde c_s^2 + \frac{2\beta_bA_b \beta_c A_c}{1+\frac{m^2a^2}{k^2}} \frac{R}{R+1})\frac{\Phi}{3\tilde c_s^2}.
\label{the}
\end{equation}
This can also be written
\begin{equation}
\Theta= \frac{\tilde \delta}{3 (1+R)^{1/2}} +\frac{R}{R+1}(1+\frac{9\Omega_b {\cal H}^2 }{{k^2}}\tilde \beta_b^2 + 2 \tilde\beta_b \tilde \beta_c )\frac{\Phi}{3\tilde c_s^2}
\label{they}
\end{equation}
where we have defined the effective couplings
\begin{equation}
\tilde \beta_{b,c}= \frac{\beta_{b,c} A_{b,c}}{1+ \frac{m^2a^2}{k^2}}.
\end{equation}
There are three sources of alteration to the peak structure. First of all the speed of sound is modified due to $\tilde \beta_b$. Then the growth of Newton's potential is also different from the one in General Relativity due to $\tilde\beta_c$. Finally, the amplitude of the term in $\Phi$ appearing in the expression for $\Theta$ is also modified with a contribution depending on $\tilde\beta_b\tilde\beta_c$.

\subsection{Phenomenology}

Assuming that the equation of state varies abruptly at equality we can write
\begin{equation}
a=a_{eq}(\frac{1+3w}{3(1+w)} \frac{\eta -\eta_{eq}}{(1+z_{eq}) t_{eq}} +1)^{2/(1+3w)}
\end{equation}
where $\eta$ is the conformal time.
Defining
\begin{equation}
k_{eq}(w)= \frac{2}{3(1+w)} \frac{1}{(1+z_{eq})t_{eq}}
\end{equation}
in both the matter and radiation eras,
we find that scales enter the horizon when ${\cal H}_k=k$ at
\begin{equation}
\frac{a_k}{a_{eq}} = (\frac{k_{eq}(w)}{k})^{2/3(1+w)}
\end{equation}
corresponding to
\begin{equation}
\eta_k= \eta_{eq}+ \frac{2}{1+3w}(\frac{1}{k}- \frac{1}{k_{eq}(w)}).
\end{equation}
Scales with $k>k_{eq}(w)$ enter the horizon before equality at
\begin{equation}
\eta_k= \eta_{eq}+(\frac{1}{k}- \frac{1}{k_{eq}(w)})
\end{equation}
We consider scales entering the horizon in the radiation era and put $k_{eq}= k_{eq}(1/3)$, then we have
\begin{equation}
k_{eq}= \frac{1}{2} \frac{1}{(1+z_{eq})t_{eq}}
\end{equation}
where
\begin{equation}
t_{eq}= \frac{t_0}{(1+z_{eq})^{3/2}}
\end{equation}
Scales corresponding to  the CMB peaks are given by
\begin{equation}
k_p= p \frac{\pi}{r_s(\eta_{LS})}
\end{equation}
where a good approximation is
\begin{equation}
r_s(\eta_{LS})\approx \frac{1}{3} (\eta_{LS} - \eta_k)
\end{equation}
or equivalently
\begin{equation}
r_s(\eta_{LS})\approx \frac{1}{3} (\eta_{LS} - \eta_{eq} + \frac{1}{3} (\frac{1}{k} -\frac{1}{k_{eq}}))
\end{equation}
and
\begin{equation}
\eta_{LS} - \eta_{eq}= \frac{2}{k_{eq}} ((\frac{1+z_{eq}}{1+ z_{LS}})^{1/2}-1).
\end{equation}
The corresponding Compton wavelength is
\begin{equation}
m_p=(1+z_{LS}) k_p.
\end{equation}
It is convenient to compare this to the Hubble rate (neglecting the effect of the late acceleration of the universe)
\begin{equation}
H_{LS}= (1+z_{LS})^{3/2} H_0
\end{equation}
so
\begin{equation}
\frac{m^2_p}{H^2_{LS}}= (1+z_{LS})^{-1} \frac{k_p^2}{H_0^2}.
\end{equation}
Numerically we find with these approximations that $k_{1}\approx 0.035 (\rm{h^{-1}.Mpc})^{-1}$ and
\begin{equation}
\frac{m^2_p}{H^2_{LS}}\approx 12.8 p^2.
\end{equation}
This is not satisfied for dilatons as $m^2/H^2 =O(A_2)$ is very large \cite{Brax:2010gi}.
As $\beta$ is small for dilatons, there should be no CMB effect in this class of models.

In the following, we set the initial conditions for a mode $k$ at $\eta_k$.  In the figures, we have selected wavenumbers corresponding to the first two peaks and chosen a power law dependence for the mass of the scalar field
\begin{equation}
m(a)= m_c (\frac{a}{a_{LS}})^{r}
\end{equation}
where $m_c$ is a characteristic scale depending on the  model. With this parametrisation, the mass is always greater than the Hubble rate since equality provided
\begin{equation}
r+\frac{3}{2} < \frac{\ln\frac{m_c}{H_{LS}}}{\ln\frac{1+z_{eq}}{1+z_{LS}}}.
\end{equation}
Scales outside the Compton radius enter the Compton radius when
\begin{equation}
k= \tilde a_k m(\tilde a_k)
\end{equation}
after horizon entry.
This implies that for a given $k$, gravity is modified as soon as
$m(a) a\le k$. We find that the Newtonian potential grows logarithmically in the radiation era leading to  a small growth of $k^2 \Phi$ when no modification of gravity is present. When $\beta_c>0$, Newton's potential shows an anomalous growth with $k$ as soon as small scales enter the Compton radius.
We also find that the speed of sound is largely affected by $\beta_b$ implying a shift in the peak positions.

\begin{figure}
\begin{center}
\includegraphics[width=7cm]{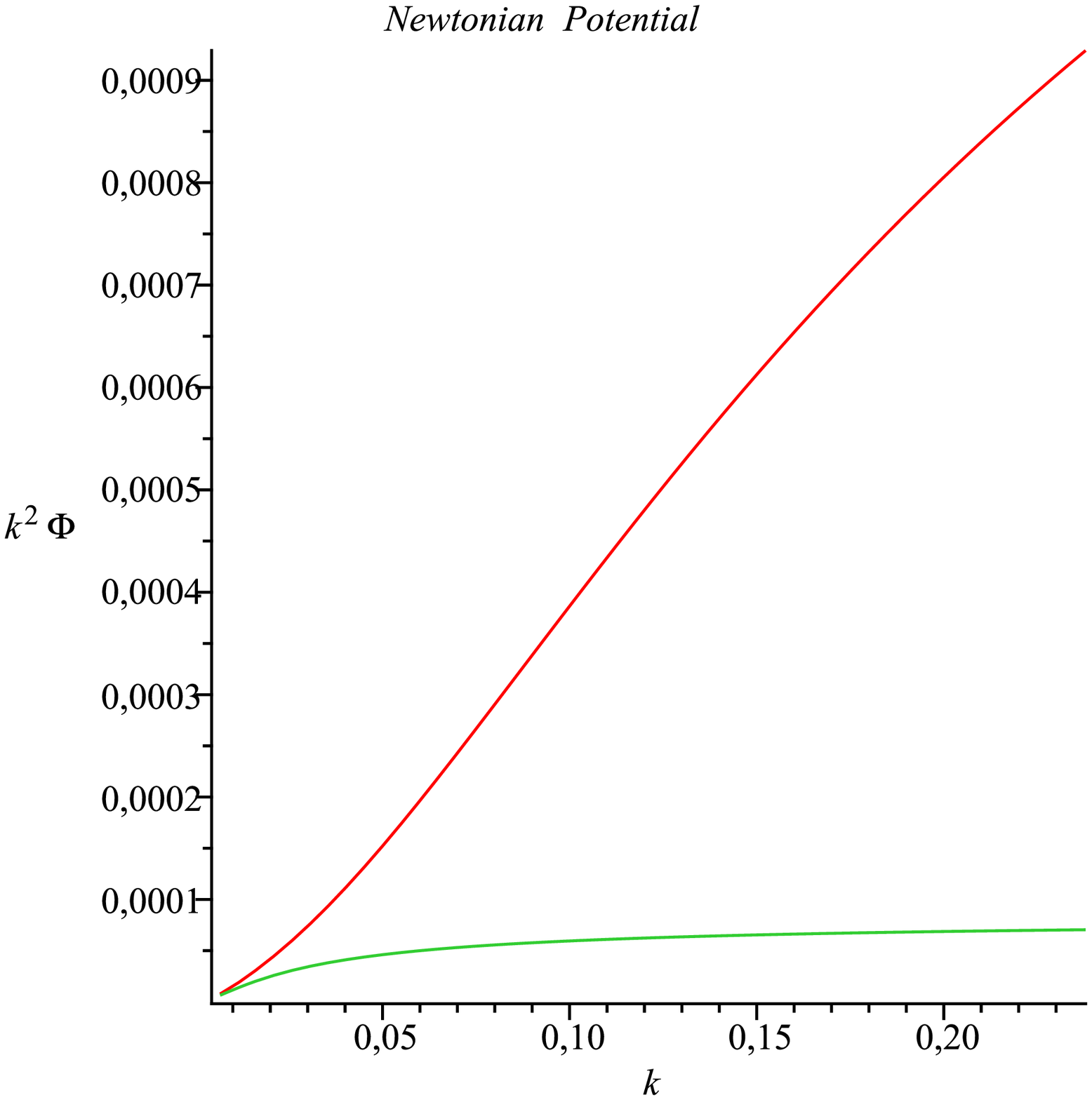}\hspace*{1cm}
\includegraphics[width=7cm]{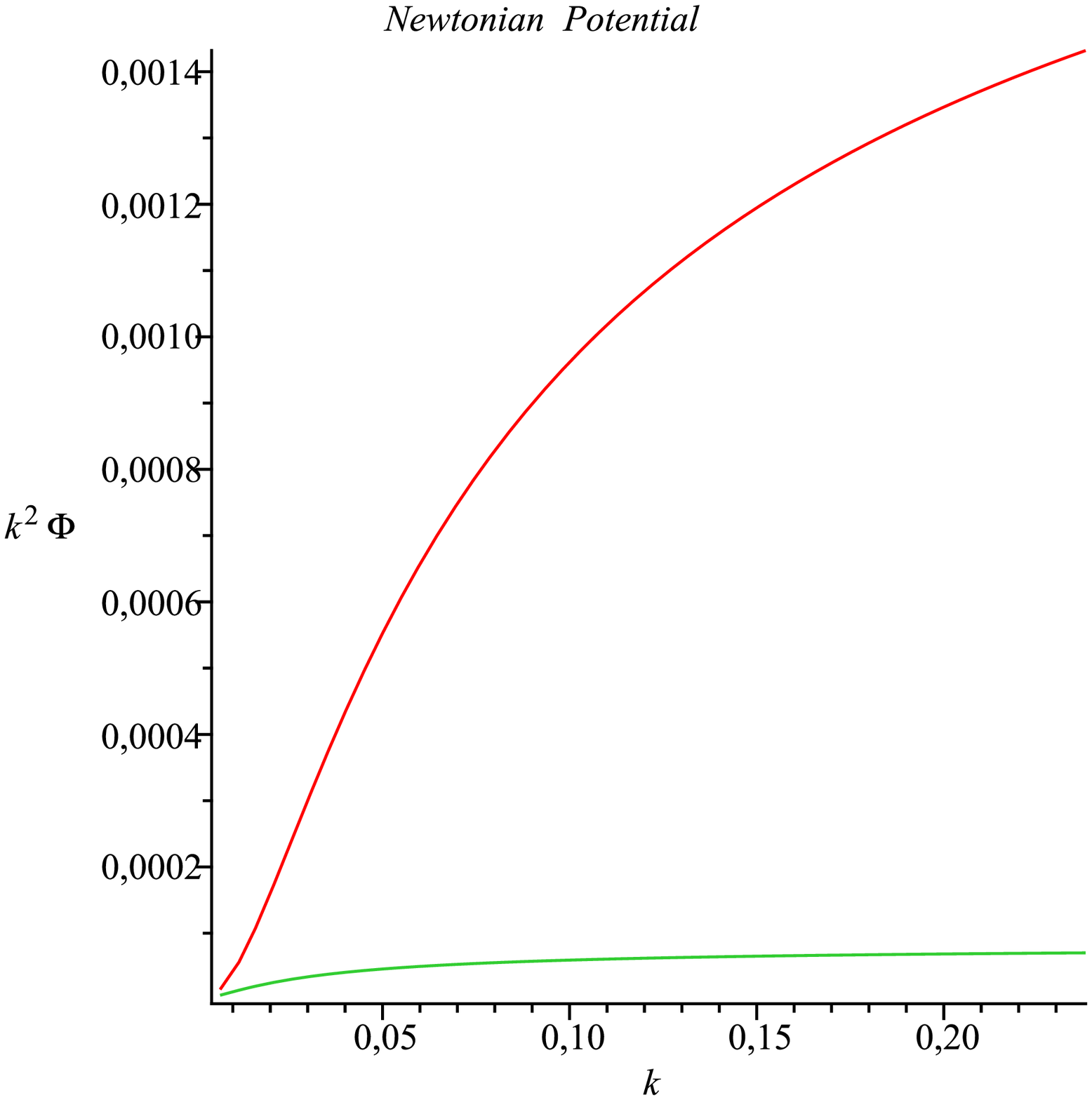}

\caption{Newton's potential $k^2\Phi$ at last scattering for $\beta_b=0$,  $\beta_c=100$,  $r=-2$ (left) and $r=1$ (right) as a function of $k$ in $({\rm h^{-1}.Mpc})^{-1}$ compared to the case with no modification of gravity. The logarithmic increase is largely modified due to the large value  of $\beta_c$. Here we have $k_1\sim 0.035 ({\rm h^{-1}.Mpc})^{-1}$, and for $r-2$, $k_c\sim 0.1 ({\rm h^{-1}.Mpc})^{-1}$,  while for $r=1$, $k_c\sim 0.006 ({\rm h^{-1}.Mpc})^{-1}$. In both cases, the anomalous growth of perturbations due to modified gravity can be seen for scales between $k_1$ and $k_c$, respectively $k_c$ and $k_1$. For smaller scales, the growth is only logarithmic.
}
\end{center}
\end{figure}

\begin{figure}
\begin{center}
\includegraphics[width=7cm]{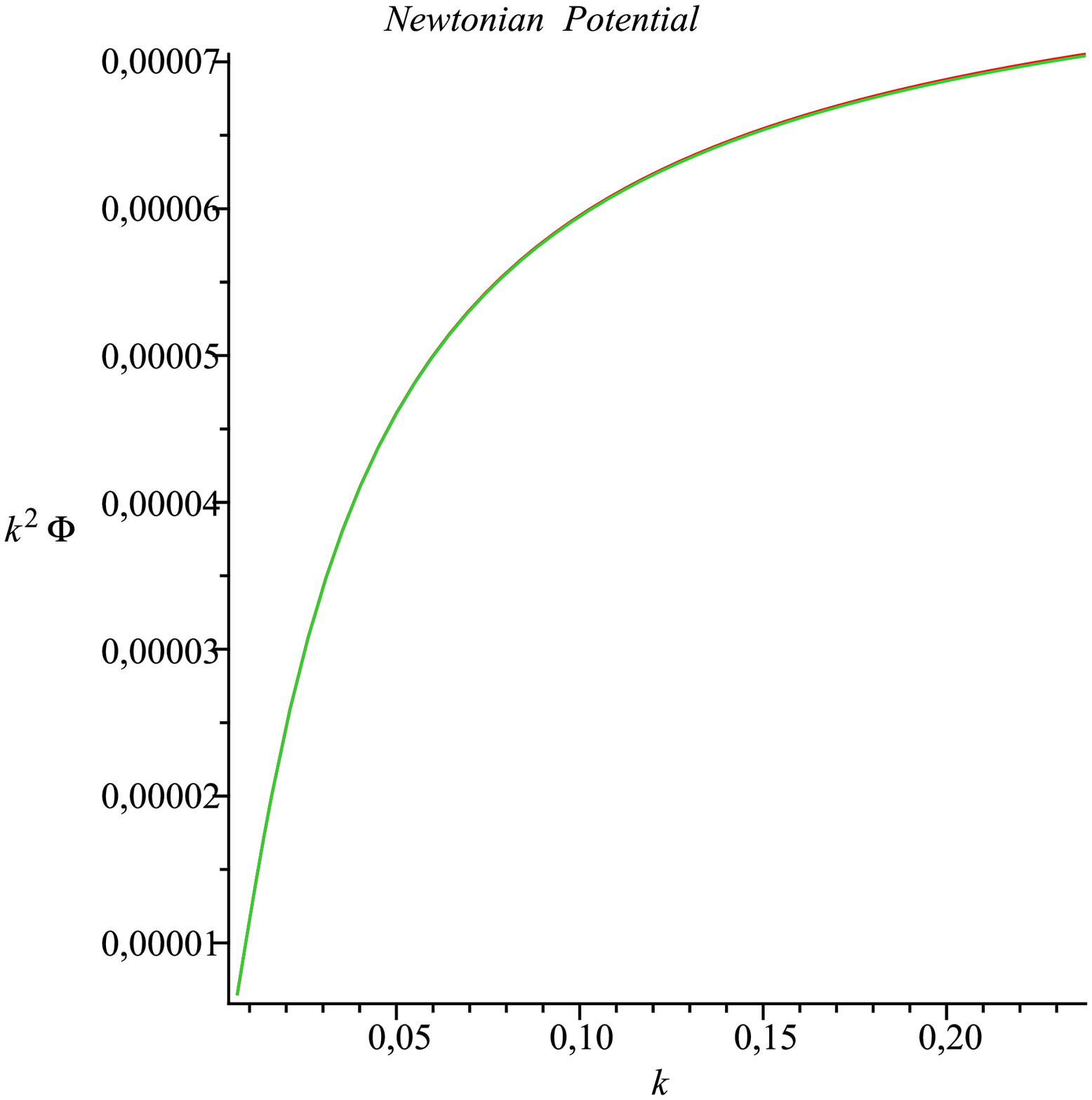}\hspace*{1cm}
\includegraphics[width=7cm]{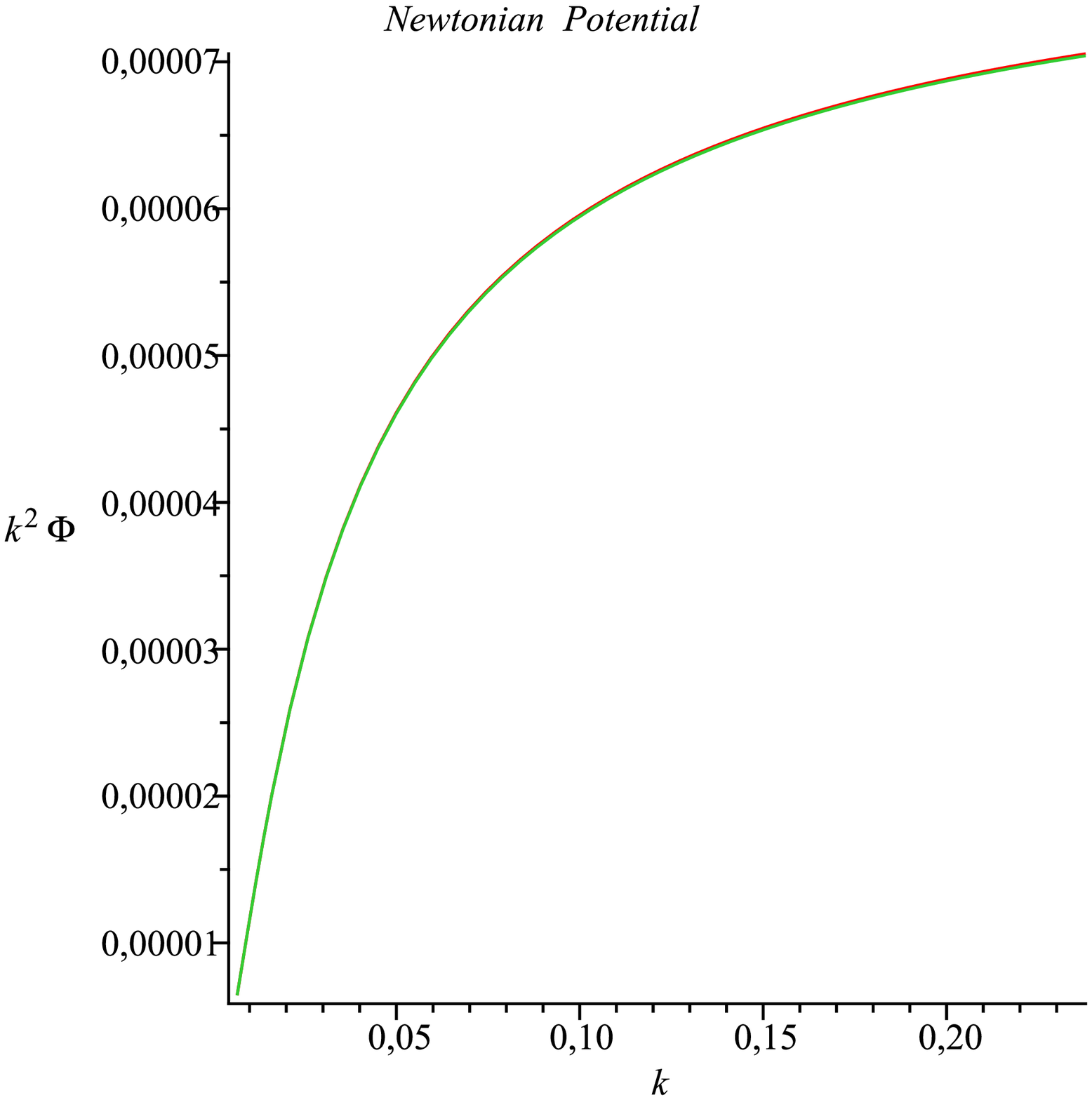}

\caption{Newton's potential $k^2\Phi$ at last scattering for $\beta_b=2$,  $\beta_c=2$, $r=-2$ and $r=1$ as a function of $k$ in $({\rm h^{-1}.Mpc})^{-1}$ compared to the case with no modification of gravity.
}
\end{center}
\end{figure}

\begin{figure}
\begin{center}
\includegraphics[width=7cm]{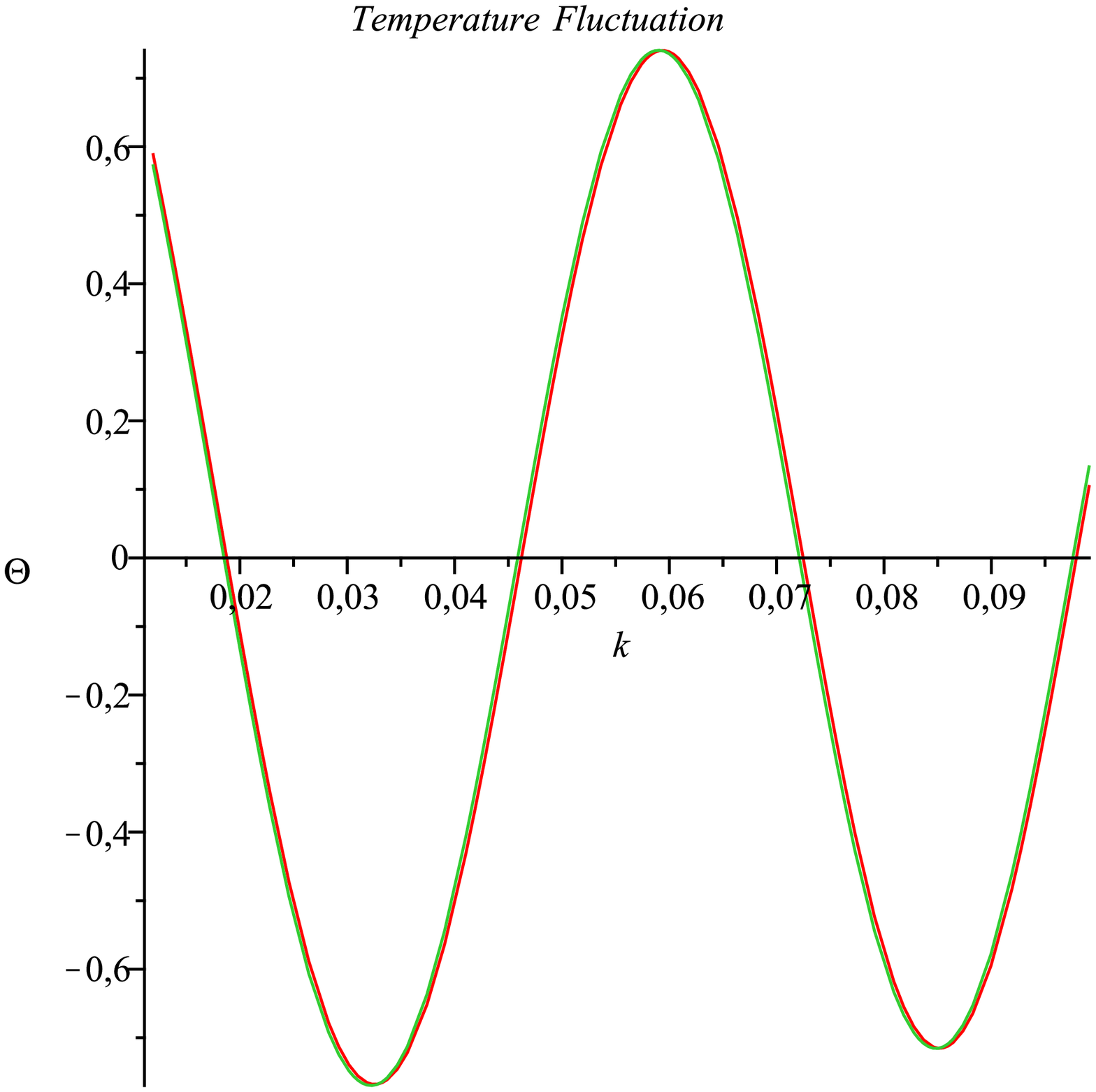}\hspace*{1cm}
\includegraphics[width=7cm]{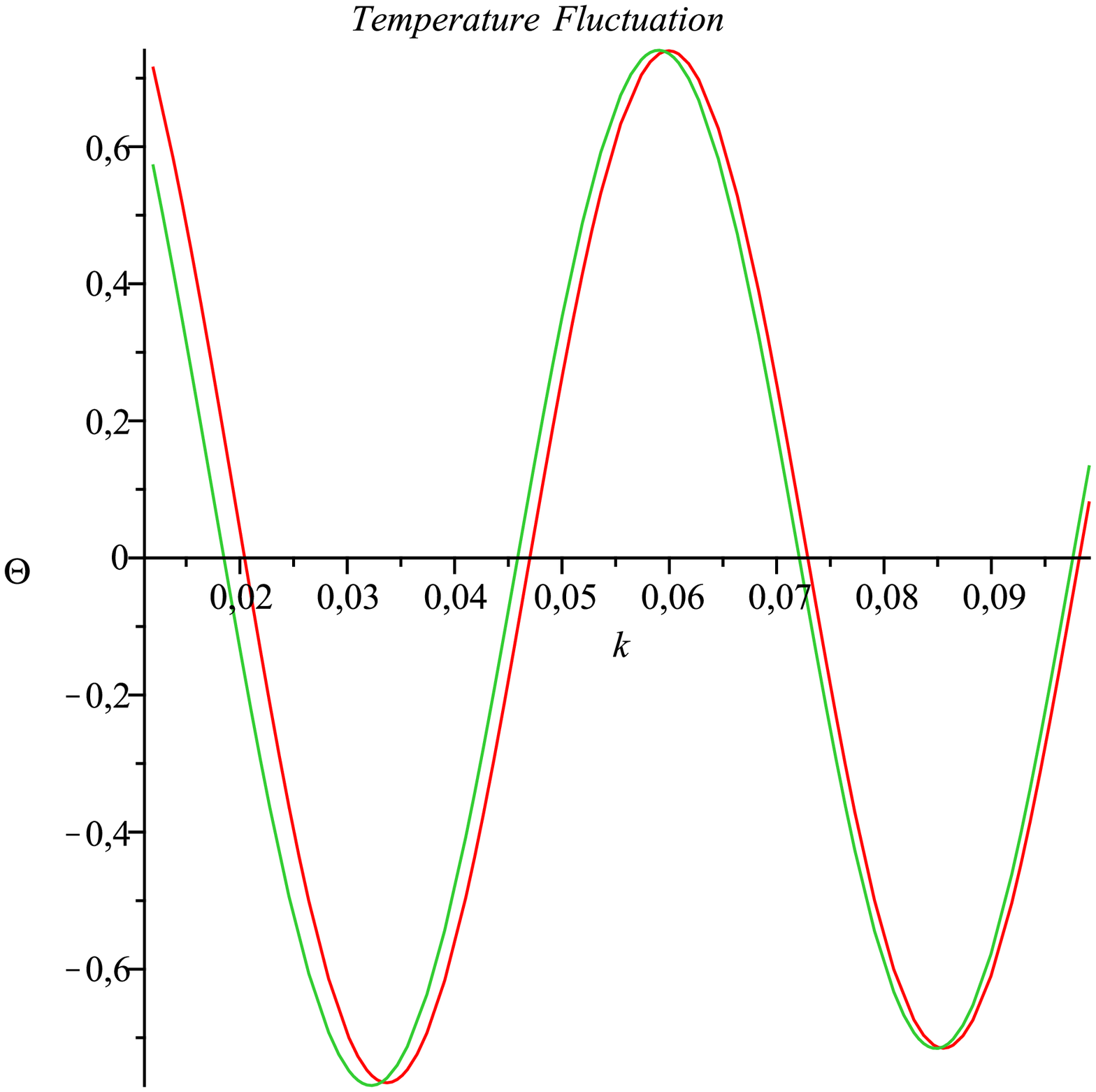}

\caption{The Sachs-Wolfe temperature fluctuation $\Theta$ at last scattering for $\beta_b=3$,  $\beta_c=0$, $r=-2$ and $r=1$ as a function of $k$ in $({\rm h^{-1}.Mpc})^{-1}$ compared to the case with no modification of gravity. A shift in the position of the peaks due to a modification of the sound speed is noticeable.}

\end{center}
\end{figure}

\begin{figure}
\begin{center}
\includegraphics[width=7cm]{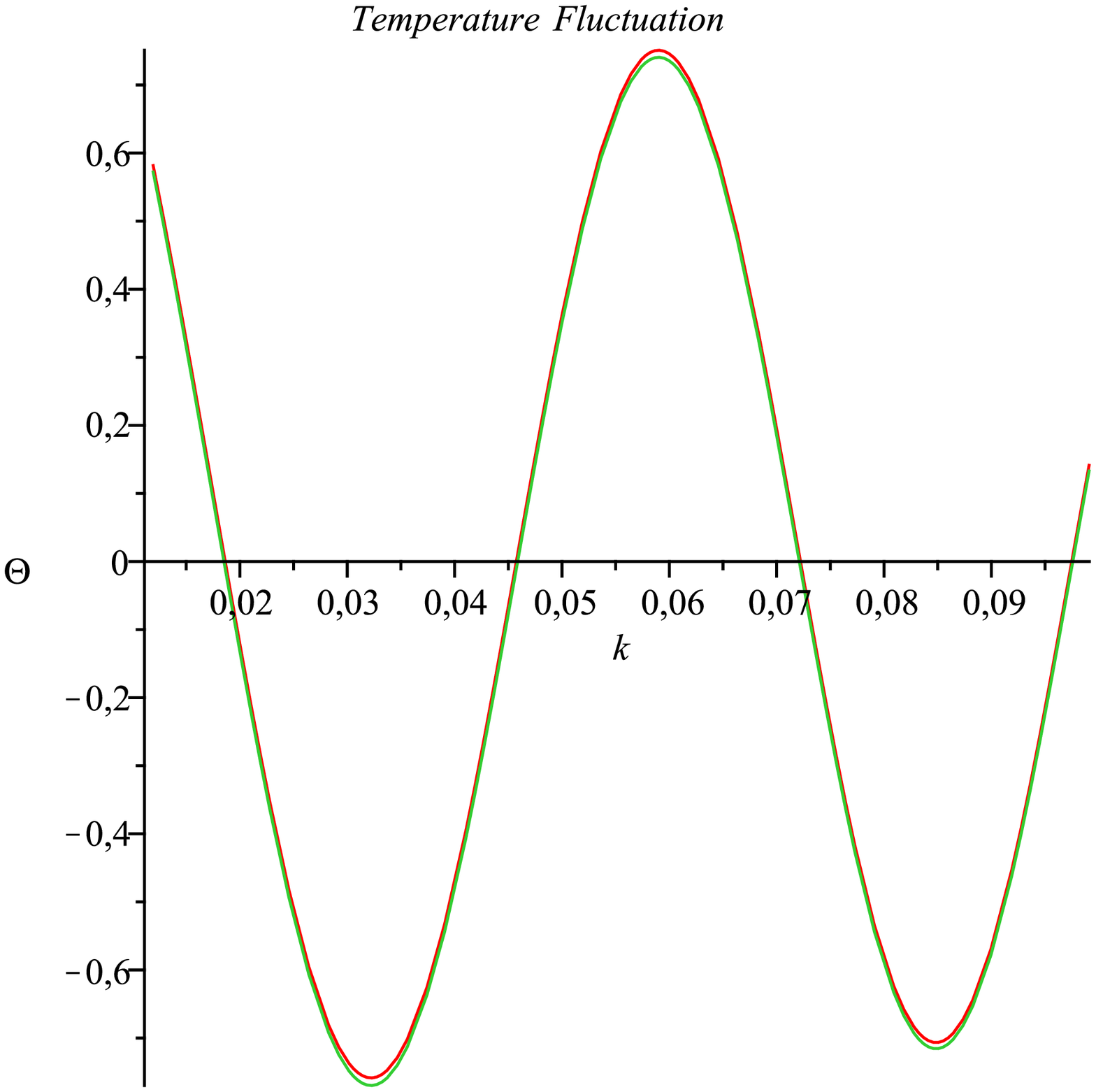}\hspace*{1cm}
\includegraphics[width=7cm]{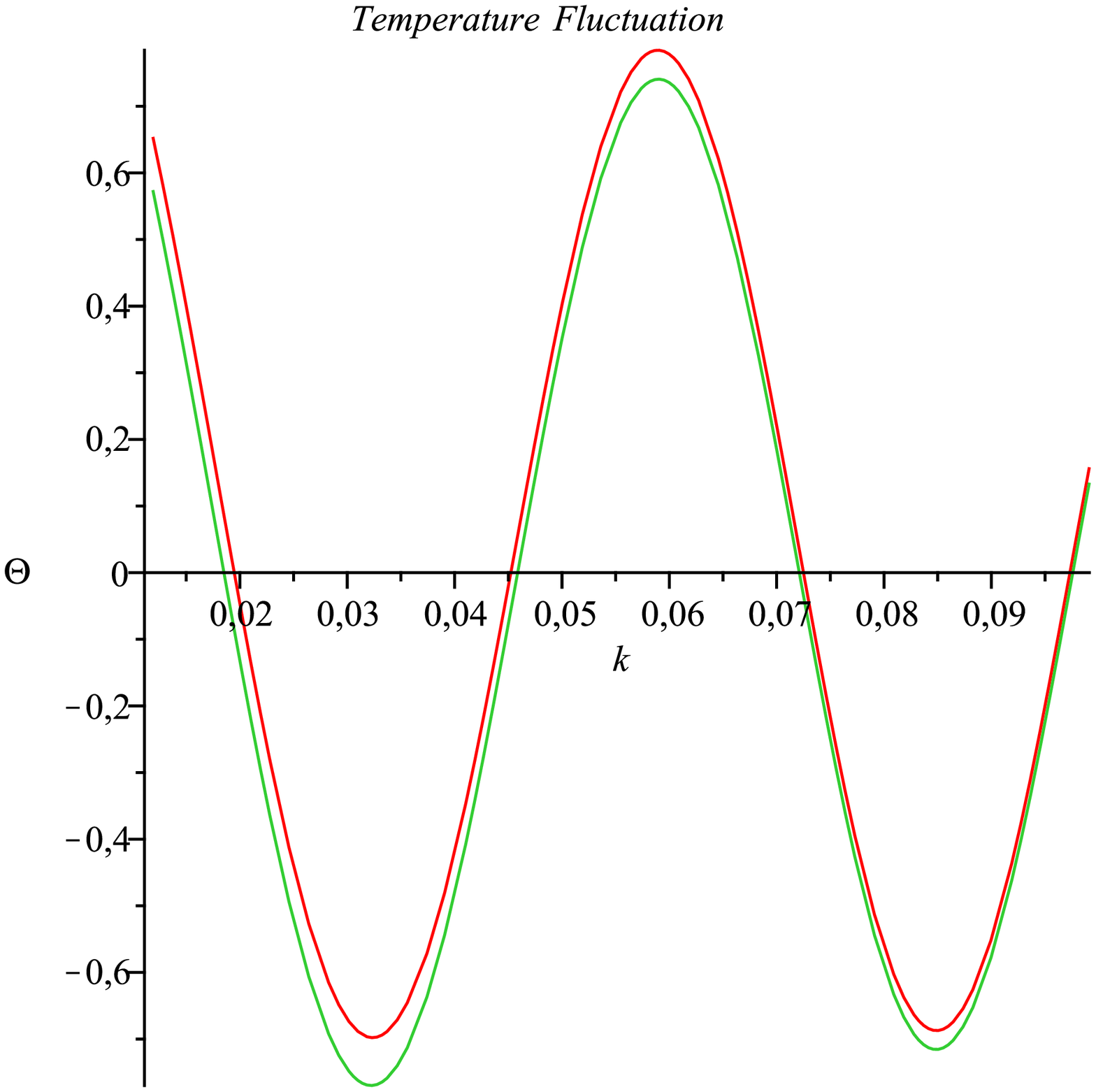}

\caption{The Sachs-Wolfe temperature fluctuation $\Theta$  at last scattering for $\beta_b=0$,  $\beta_c=100$, $r=-2$ and $r=1$ as a function of $k$ in $({\rm h^{-1}.Mpc})^{-1}$ compared to the case with no modification of gravity.
}
\end{center}
\end{figure}

\begin{figure}
\begin{center}
\includegraphics[width=7cm]{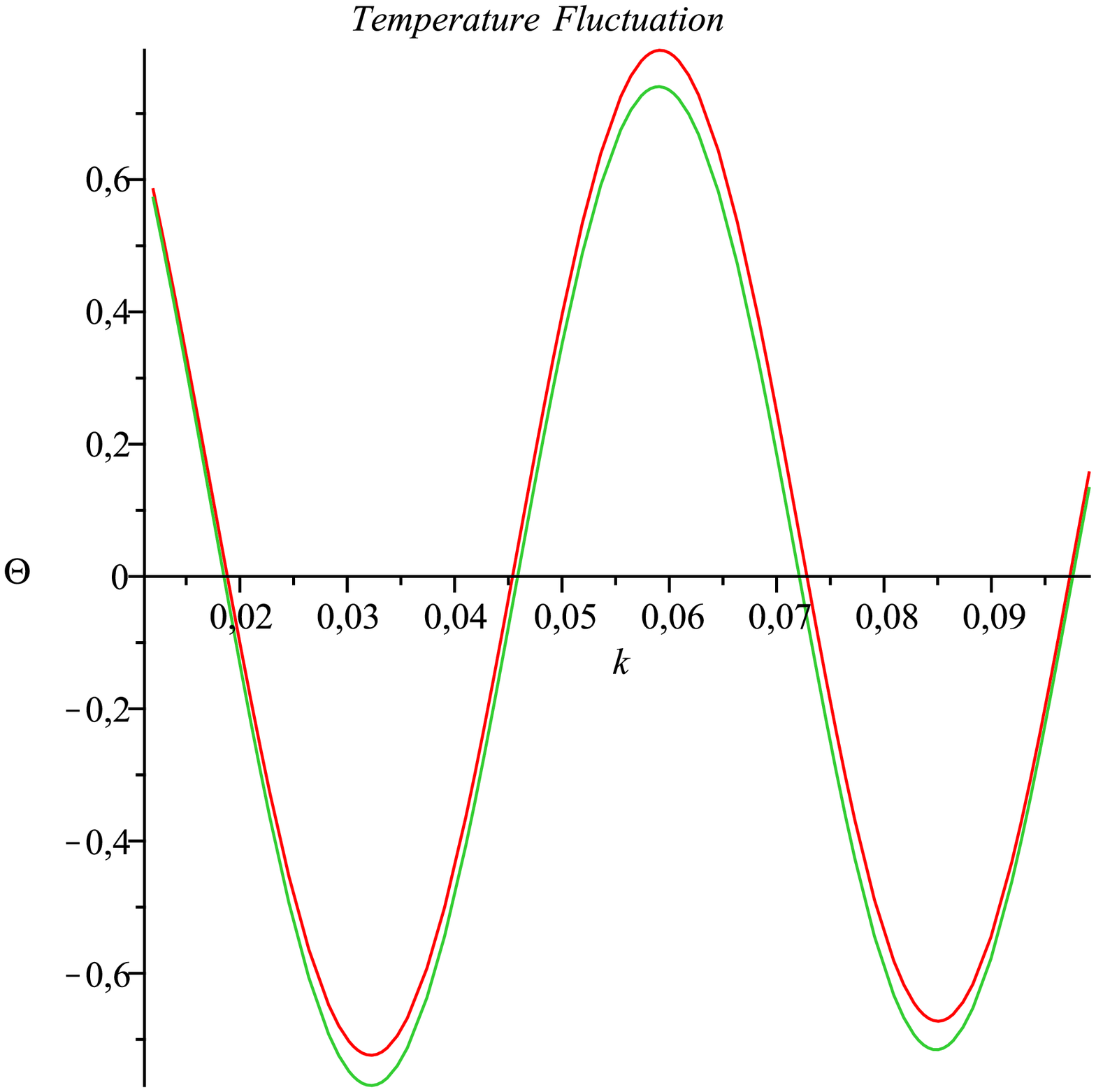}\hspace*{1cm}
\includegraphics[width=7cm]{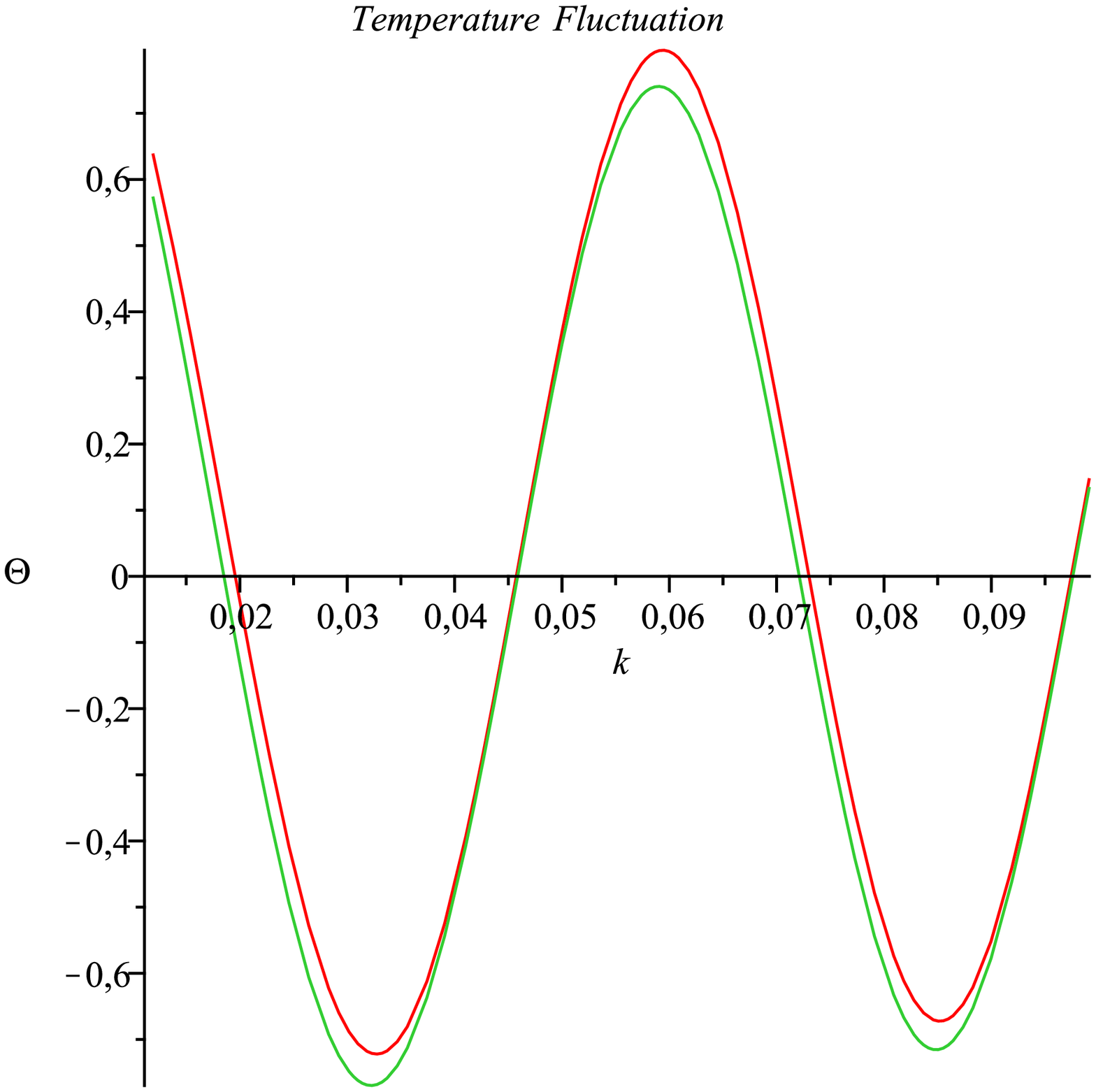}

\caption{The Sachs-Wolfe temperature fluctuation $\Theta$  at last scattering for $\beta_b=2$,  $\beta_c=2$, $r=-2$ and $r=1$ as a function of $k$ in $({\rm h^{-1}.Mpc})^{-1}$ compared to the case with no modification of gravity.
}
\end{center}
\end{figure}

The effect can be seen in the figures 1-5 where three types of couplings have been chosen. When $\beta_b$ is large and $\beta_c$ is small, here taken to be $\beta_b=3$ and  $\beta_c=0$, we find that the peaks are shifted due to a change of the sound horizon. On the other hand, when $\beta_c$ is large and $\beta_b$ is small, here $\beta_c=100$ and $\beta_b=0$,  the amplitude of the peaks is increased. When both are of the same order, the effects are combined together with a new effect coming from
\begin{equation}
\delta\Theta=\tilde\beta_b \tilde \beta_c \frac{2 R}{R+1} \frac{\Phi}{3\tilde c_s^2}
\label{corr}
\end{equation}
whose influence can be seen even when $\beta_{b,c}$ are small and no direct deviation on the speed of sound and the amplitude of Newton's potential is noticeable.
In figure 2 and 5 with $\beta_c=\beta_b=2$, we show how little effect modified gravity has on Newton's potential and the large effect on the CMB peaks from (\ref{corr}).

 In all cases we have chosen $m_c=m_1$ to enhance the modification of the CMB spectrum. In this case we must impose that $r<1.2$ and we have chosen two relevant cases with $r=1$ and $r=-2$ where $am(a)$ is either increasing or decreasing respectively.
When $am(a)$ increases, scales leave the Compton radius eventually. When $am(a)$ decreases, scales enter the Compton radius.
In the first case, for large enough $k$, scales are in the Compton radius at horizon entry and will eventually leave it. In the latter,
scales which are not within the Compton radius at horizon entry will eventually enter the Compton radius. This results in different behaviours for Newton's potential as a function of $k$. When $\beta_c$ is large enough and $am(a)$ increases, scales with large $k$ tend to increase with a modified rate compared to General Relativity for a longer period of time. The charactistic scale $k_c$ is $k_c=0.006 h\cdot {\rm Mpc}^{-1}$ when $r=1$ and $k_c=0.1 h\cdot {\rm Mpc}^{-1}$ when $r=-2$. In figure 1, one can see the power law growth between $k_1$ and $k_c$ when $r=-2$, respectively $k_c$ and $k_1$ for $r=1$, followed by a logarithmic rise for larger values of $k$.

These examples have been chosen to accentuate the effects. The couplings $\beta_{b,c}$ have been taken to be voluntarily too large. We expect that the tight bounds from the Planck satellite would lead to stringent constraints on the models.

\subsection{Engineering modified gravity}

In this section we will show that one can always engineer any modified gravity model characterised by the couplings to matter $\beta_{b,c}$ and a mass function $m(a)$. This can be achieved using
chameleon theories in the cosmological regime which are determined by the couplings $\beta_{b,c}$ and the time dependent mass $m(a)$ as a function of the scale factor,
as long as $m \gg H$ implying that the minimum of the effective potential $V_{\rm eff}$ is a dynamical attractor. Indeed, the mass at the minimum of the effective potential leads to the constraints
\begin{equation}
V''\equiv \frac{d^2V}{d\phi^2}= m^2 (a) - \sum_i \beta_i ^2 A_i \frac{\rho_i}{m_{\rm Pl}^2}-\sum_i \beta_i' A_i \frac{\rho_i}{m_{\rm Pl}}
\end{equation}
where the couplings $\beta_i$ can be field dependent.
Using the minimum equation
\begin{equation}
\frac{dV}{d\phi}=-\sum_i \beta_i A_i \frac{\rho_i}{m_{\rm Pl}}
\end{equation}
we deduce that the field evolves according to
\begin{equation}
\frac{d\phi}{dt}=\frac{3H}{m^2} \sum_i \beta_i A_i \frac{\rho_i}{m_{\rm Pl}}.
\end{equation}
To simplify the analysis, let us
assume that $A_i=\exp(\kappa_4 \beta_i \phi)$ and $\kappa_4\phi \ll 1$, we obtain that
\begin{equation}
\phi(a)=  \frac{3\beta}{m_{\rm Pl}}\int_{a_{\rm ini}}^a \frac{1}{a m^2(a)}\rho (a)  da +\phi_0
\end{equation}
where the sum is over non-relativistic species and we have defined
an effective coupling constant $\beta$ as
\begin{equation}
\beta \rho= \sum \beta_i \rho_i
\end{equation}
and $\rho=\sum_i \rho_i$.
The minimum equation implies that
\begin{equation}
V=V_0 -3\beta^2  \int_{a_{\rm ini}}^a \frac{1}{am^2(a)} \frac{\rho^2}{m^2_{\rm Pl}} da.
\end{equation}
This defines the potential parametrically. Of course, one should check that local tests of gravity are then satisfied with this potential.

\subsection{Power law potentials}

Let us give an explicit example:
\begin{equation}
m(a)= m_{\rm ini}(\frac{a}{a_{\rm ini}})^r.
\label{po}
\end{equation}
In the matter dominated era, we have
$H(a)= H_{\rm ini} (\frac{a_{\rm ini}}{a})^{3/2}$ and  $m_{\rm ini}\gg H_{\rm ini}$.
We then obtain
\begin{equation}
\phi (a)= \phi_o + \frac{3\beta \rho_{\rm ini}}{m_{\rm ini}^2 m_{\rm Pl}} \frac{1}{3+2r}\left ( 1- (\frac{a_{\rm ini}}{a})^{3+2r}\right ).
\end{equation}
As long as $3+2r<0$, we can choose $\phi_0$ such that
\begin{equation}
\phi= \frac{3\beta \rho_{\rm ini}}{m_{\rm ini}^2 m_{\rm Pl}} \frac{1}{\vert 3+2r\vert} (\frac{a_{\rm ini}}{a})^{3+2r}
\end{equation}
This is a power law behaviour.
Similarly we obtain
\begin{equation}
V= \Lambda_0^4 + \frac{3\beta^2 \rho_{\rm ini}^2}{m_{\rm ini}^2 m^2_{\rm Pl}}\frac{1}{6+2r} (\frac{a_{\rm ini}}{a})^{6+2r}
\end{equation}
where $\Lambda_0$ is the vacuum energy.
This leads to a usual chameleon potential where we can identify
\begin{equation}
n=-\frac{6+2r}{3+2r}
\end{equation}
and
\begin{equation}
\Lambda^{4+n}=\frac{3\beta^2 \rho_{\rm ini}^2}{m_{\rm ini}^2 m^2_{\rm Pl}}\frac{1}{6+2r}(\frac{3\beta \rho_{\rm ini}}{m_{\rm ini}^2 m_{\rm Pl}} \frac{1}{\vert 3+2r\vert})^n
\end{equation}
where $-3<r<-3/2$.
This defines a chameleonic model coupled to both baryons and CDM particles.
When $r$ is greater than $-3/2 $, the models are also of the power law type with positive powers of the field.

Scales such as $m_{\rm ini}^{-1}$ close to the CMB peak scales and the choice of a power law evolution of the mass function (\ref{po}) between the last scattering surface and now may lead to an incompatibility of the models with both the recent growth of large scale structures and/or local gravity tests in the solar system and the laboratory \cite{new}. Models where effects of modified gravity could be present both in the recent past of the Universe and at the last scattering surface are more likely to require engineering the function $m(a)$ around $a\sim 10^{-3}$ and $a\sim 1$ with different functional forms, for instance different power law functions, and interpolating between these two regimes where constraints are scarce. This is left for future work

\subsection{Discussion}

We have shown that for large enough couplings to the baryons and/or CDM, effects on the CMB peaks can be induced provided the mass of the
scalar field mediating modified gravity is tuned to be close to the scales of the CMB peaks.
We also know that strong constraints on modified gravity can be obtained at low redshift using large scale structures of the Universe \cite{Brax:2009ab}.
The scales corresponding to the CMB and LSS are different and can be connected by an appropriate choice of the mass function $m(a)$. Hence,
the CMB and LSS constraints can be seen to be independent as they rely on two different ranges of the scale factor for the function $m(a)$. Consequently,
unless one can infer a function $m(a)$ from an underlying theory, one must take the constraints given by the position of  the CMB peaks  and the LSS constraints on the growth of structures as independent snapshots of an unknown function $m(a)$. In fact, at the linear level, modified gravity models are characterised by their couplings
to matter and the mass function $m(a)$.

In this paper we have focused on the CMB physics and its modification due to a scalar field with different couplings to baryons and CDM. We have seen that the effect
of the baryon coupling is prominent on the speed of sound. In particular, this coupling cannot be too large when the mass of the scalar field is of the order of the
scale of the CMB peaks. For very large couplings, the speed of sound becomes imaginary. For smaller values of the baryon coupling, the position of the CMB peaks is shifted due to the alteration of the speed of sound. The CMB coupling influences the amplitude of the Newtonian potential. The effect is more relevant when the mass scale $am(a)$ increases as short scales enter the Compton radius early and have more time to grow anomalously.

The subsequent evolution of the $m(a)$ function is not constrained by the CMB physics. It is plausible that 21 cm physics may give some indication on the shape of the function $m(a)$ between the CMB and LSS epochs.  Constraints on structure formation at low redshifts exist already while future surveys will give tighter results.
On the whole, the landscape of possible modified gravity models, characterised by the couplings to matter and the mass function,  is almost completely unchartered.
In this paper, we have suggested that CMB physics may help unraveling some part of its geography.

\section{Conclusion}

Modified gravity models involving a scalar field and generating the acceleration of the Universe seem to require a screening mechanism to
evade local gravitational tests. The same models must lead to a very slow evolution of the scalar field since BBN in order to comply
with the stringent bounds on the formation of the elements. In general, this results in the fact that these models behave like a $\Lambda$-CDM
model since BBN at the background level. At the perturbative level and on large scales, the models are characterised by their couplings to matter and
the mass function $m(a)$, i.e. the mass of the scalar field as a function of time. We have shown that given these inputs one can engineer a
modified gravity model of the chameleon type (at least when the couplings are constant). We have focussed on the CMB peaks and the effect of modified gravity on their structure. We have found that three possible effects can occur. For large enough couplings to CDM and a mass scale of order of the sound horizon at last scattering, we
have a strong modification of Newton's potential and an increase of the CMB peak amplitudes. Far large enough couplings to baryons, the speed of sound is reduced implying a shift to large values of the momenta for the CMB peaks. Finally, even for moderate couplings, the Sachs-Wolfe temperature could be altered by the combined effect of the
couplings to baryons and CDM.

We hope that these effects could be attainable with the forthcoming Planck experiment. Unfortunately, this would uncover only one corner of the
modified gravity puzzle as large scale structures  are sensitive to the mass function at much later times. Maybe 21 cm physics could help
connecting information from both epochs. This is left for future work.

\section{Acknowledgements}
We thank Baojiu Li and Anthony Challinor for discussions and
Christophe Ringeval and Julien Lesgourges for their
comments on an earlier version of the manuscript.
This work is supported in part by STFC (ACD). ACD wishes
to thank CEA Saclay for hospitality whilst this work was in progress.


\begin{thebibliography}{99}



\bibitem{Jain:2010ka}
  B.~Jain, J.~Khoury,
  Annals Phys.\  {\bf 325 } (2010)  1479-1516.
  [arXiv:1004.3294 [astro-ph.CO]].

\bibitem{Clifton:2011jh}
  T.~Clifton, P.~G.~Ferreira, A.~Padilla, C.~Skordis,

  [arXiv:1106.2476 [astro-ph.CO]].


\bibitem{Khoury:2003aq}
  J.~Khoury and A.~Weltman,
  Phys.\ Rev.\ Lett.\  {\bf 93} (2004) 171104
  [arXiv:astro-ph/0309300].
\bibitem{Khoury:2003rn}
  J.~Khoury and A.~Weltman,
  Phys.\ Rev.\  D {\bf 69}, 044026 (2004)
  [arXiv:astro-ph/0309411].

\bibitem{Brax:2004qh}
Ph.~ Brax, C.~ van de Bruck, A-C~ Davis, J.~ Khoury and  A.~Weltman,
Phys.Rev.D70:123518,2004
[arXiv: astro-ph/0408415]

\bibitem{Mota:2006ed}
  D.~F.~Mota and D.~J.~Shaw,
  Phys.\ Rev.\ Lett.\  {\bf 97} (2006) 151102
  [arXiv:hep-ph/0606204].

\bibitem{Mota:2006fz}
  D.~F.~Mota and D.~J.~Shaw,
  Phys.\ Rev.\  D {\bf 75} (2007) 063501
  [arXiv:hep-ph/0608078].



\bibitem{Hinterbichler:2010es}
  K.~Hinterbichler, J.~Khoury,
  Phys.\ Rev.\ Lett.\  {\bf 104 } (2010)  231301.
  [arXiv:1001.4525 [hep-th]].


\bibitem{Brax:2010gi}
  P.~Brax, C.~van de Bruck, A.~-C.~Davis, D.~Shaw,
  Phys.\ Rev.\  {\bf D82 } (2010)  063519.
  [arXiv:1005.3735 [astro-ph.CO]].

\bibitem{Nicolis:2008}
  A.~Nicolis, R.~Rattazzi, E.~Trincherini,
  Phys.\ Rev.\  {\bf D79 } (2009)  064036.
  [arXiv:0811.2197 [hep-th]].

\bibitem{Olive:2007aj}
  K.~A.~Olive, M.~Pospelov,
  Phys.\ Rev.\  {\bf D77 } (2008)  043524.
  [arXiv:0709.3825 [hep-ph]].

\bibitem{Chen:1999qh}
  X.~-l.~Chen, M.~Kamionkowski,
  Phys.\ Rev.\  {\bf D60 } (1999)  104036.
  [astro-ph/9905368].

\bibitem{Acquaviva:2004ti}
  V.~Acquaviva, C.~Baccigalupi, S.~M.~Leach, A.~R.~Liddle, F.~Perrotta,
  Phys.\ Rev.\  {\bf D71}, 104025 (2005).
  [astro-ph/0412052].


\bibitem{Ali:2010gr}
  A.~Ali, R.~Gannouji, M.~Sami,
  Phys.\ Rev.\  {\bf D82 } (2010)  103015.
  [arXiv:1008.1588 [astro-ph.CO]].


\bibitem{Bertotti:2003rm}
  B.~Bertotti, L.~Iess, P.~Tortora,
  Nature {\bf 425 } (2003)  374.



\bibitem{Bean:2007ny}
  R.~Bean, E.~E.~Flanagan, M.~Trodden,
  Phys.\ Rev.\  {\bf D78}, 023009 (2008).
  [arXiv:0709.1128 [astro-ph]].

\bibitem{Brax:2008hh}
  P.~Brax, C.~van de Bruck, A.~-C.~Davis, D.~J.~Shaw,
  Phys.\ Rev.\  {\bf D78 } (2008)  104021.
  [arXiv:0806.3415 [astro-ph]].

\bibitem{Rhodes:2003ev}
  C.~S.~Rhodes, C.~van de Bruck, P.~.Brax, A.~C.~Davis,
  Phys.\ Rev.\  {\bf D68}, 083511 (2003).
  [astro-ph/0306343].


\bibitem{Mainini:2010ng}
  R.~Mainini, D.~F.~Mota,

  [arXiv:1011.0083 [astro-ph.CO]].


\bibitem{Hu:2007nk}
  W.~Hu, I.~Sawicki,
  Phys.\ Rev.\  {\bf D76 } (2007)  064004.
  [arXiv:0705.1158 [astro-ph]].


\bibitem{Hu:1994uz}
  W.~Hu, N.~Sugiyama,
  Astrophys.\ J.\  {\bf 444 } (1995)  489-506.
  [astro-ph/9407093].

  \bibitem{new} P.~Brax, A.~-C.~Davis and B.~ Li, in preparation.

\bibitem{Brax:2009ab}
  P.~Brax, C.~van de Bruck, A.~-C.~Davis, D.~Shaw,
  JCAP {\bf 1004 } (2010)  032.
  [arXiv:0912.0462 [astro-ph.CO]].



\end{thebibliography}
\end{document}